\documentclass[
showpacs,preprintnumbers,amsmath,amssymb,showkeys]{revtex4}
\usepackage{amssymb}

\usepackage{graphicx}
\usepackage{dcolumn}
\usepackage{bm}

\def\nn{\noindent}
\newcommand{\be}{\begin{equation}}
\newcommand{\ee}{\end{equation}}
\newcommand{\bea}{\begin{eqnarray}}
\newcommand{\eea}{\end{eqnarray}}
\def\Re{{\cal R \mskip-4mu \lower.1ex \hbox{\it e}\,}}
\def\Im{{\cal I \mskip-5mu \lower.1ex \hbox{\it m}\,}}

\def\tev{\,{\ifmmode\mathrm {TeV}\else TeV\fi}}
\def\gev{\,{\ifmmode\mathrm {GeV}\else GeV\fi}}
\def\mev{\,{\ifmmode\mathrm {MeV}\else MeV\fi}}
\def\to{\rightarrow}
\def\pslash{p\,\!\!\!\!\slash }
\def\qslash{q\!\!\!\slash }
\def\kslash{k\!\!\!\slash }

\begin{document}
\title{Noncommutative QED corrections to process $e^+e^-\to\mu^+\mu^-\gamma$
 at linear collider energies}
\author{Yongming Fu, Zheng-Mao Sheng\thanks{Corresponding author.E-mail:zmsheng@css.zju.edu.cn}
 \\
\footnotesize{\it Department of Physics, Zhejiang University,
Hangzhou 310027, China
 }}

\begin{abstract}
The cross section for process $e^+e^-\to\mu^+\mu^-\gamma$ in the
framework of noncommutative quantum electrodynamics(NC QED) is
studied. It is shown that the NC correction of scattering sections
is not monotonous enhancement with total energy of colliding
electrons, but there is an optimal collision energy to get the
greatest NC correction. Moreover, there is a linear relation between
NC QED scale energy and the optimal collision energy.  The
experimental methods to improve the precision of determining NC
effects are discussed, because this process is  an ${\cal
O}(\alpha^3) $  NC QED process, high precision tests are necessary.
\end{abstract}

\maketitle

\section{INTRODUCTION}
The noncommutativity of space-time was firstly introduced by
Snyder\cite{Snyder47}. It has been revived recently with the
emergence of noncommutative quantum field theories (NC QFT) from
string theories \cite{Connes98,Douglas98,SW99}, which has
attracted much attention
\cite{Sheikh-Jabbari99,Martin99,Ishibashi00,Ya00,Petriello01,Carroll01,Seiberg00,Anisimov02,Chaichian05}.
It was generally assumed that predictions of string theory and NC
effects can only be examined at Plank scale or the grand
unification scale. However, given the possibility
\cite{Witten96,Hewett01} that the onset of string effects is at
the \tev~ scale, and that gravity becomes strong at a few \tev, it
is feasible that NC effects could also show up at a few \tev.

In
Refs.\cite{Hewett01,Kamoshita,Arfaei00,Baek01,Godfrey02,Abbiendi03,sheng,Alberto05},
several NC Quantum Electrodynamics (QED) processes in $e^{+}e^{-}$
collisions have been studied. However, the NC radiative effects in
electron-positron annihilations have not been studied before. It
is necessary to include the radiative effects in considering the
NC effects for those reactions, because the radiative corrections
due to emission of hard photons become more and more important as
$e^{+}e^{-}$ collision energy increases\cite{Nuovo,petriello03}. A
linear relation between the fundamental NC scale $\Lambda_{NC}$
and the optimal collision energy at which NC correction of
scattering section arrives at the maximum, for M\"{o}ller
scattering and Bhabha Scattering was found in Ref.\cite{sheng}. It
is interesting to explore the availability of this linear relation
for reaction $e^+e^-\to\mu^+\mu^-\gamma$, because there are three
photon self-interactions for reaction $e^+e^-\to\mu^+\mu^-\gamma$
in the NC QED, which is forbidden in conventional QED.

The essential idea of NC QED is a generalization of QED based on
the usual d-dimensional space, $R^{d}$, associated with commuting
space-time coordinates to one which is based on non-commuting,
$R^{d}_{\theta}$. In such a space the conventional coordinates are
replaced by operators which no longer commute each other. The NC
space-time can be realized by the coordinate operators
satisfying\cite{SW99,Hewett01,Kamoshita} \be [\hat{X}_\mu,\hat{X}
_\nu]=i\theta_{\mu\nu}=\frac{i c_{\mu\nu}}{\Lambda_{NC}^2},
\label{NCST} \ee
 where $\theta_{\mu\nu}$ is a constant antisymmetric matrix,
 having dimension of $(length)^2$ = $(mass)^{-2}$, and $c_{\mu\nu}$ is a real antisymmetric matrix,
 whose dimensionless elements are presumably of order
 unity. In the last equality of eq.(\ref{NCST}), we have parameterized the effect in terms of an overall
 scale $\Lambda_{NC}^2$, which characterizes the threshold where NC
 effects become relevant, its role can be compared to
 that of $\hbar$ in conventional Quantum Mechanics, which represents
 the level of non-commutativity between coordinates and momenta.
The existence of a finite $\Lambda_{NC}$ represents existence of a
fundamental space-time distance below which the space-time
coordinates become fuzzy.

When the particle energy is near or higher than $\Lambda_{NC}$, we
have to replace conventional Quantum field theories (QFT) by NC
QFT which is based on noncommutative space-time. So the
determination of $\Lambda_{NC}$ is important in NC QFT. Its lower
bound has been found to be $\Lambda_{NC}>100GeV$
 \cite{Chaichian} so that the result of Lamb shift is
 consistent with the ordinary quantum mechanics.

   NC QFT can be presented in terms of conventional commuting
   QFT through the application of  Weyl-Moyal
   correspondence\cite{Sheikh-Jabbari99}:

\begin{equation}
\hat{\Phi}(\hat{X})\longleftrightarrow \Phi(x),
\end{equation}
\begin{equation}
\hat{\Phi}_1(\hat{X})\hat{\Phi}_2(\hat{X})\longleftrightarrow
\Phi_1(x)\star\Phi_2(x) \label{Weyl},
\end{equation}
where $\Phi$ represents quantum fields with $\hat{X}$ being the
set of non-commuting coordinates and $x$ corresponding to the
commuting set, and the Moyal $\star$-product is defined by
\begin{equation}
(f\star
g)(x)=exp\left(\frac{1}{2}\theta_{\mu\nu}\partial_{x^\mu}\partial_{y^\nu}\right)f(x)g(y)|_{y=x}
\label{StarP}.
\end{equation}

Thus the Lagrangians with set of non-commuting coordinates will
correspond to those with commuting set, but the ordinary products of
all fields in the Lagrangians of their commutative counterparts
should be replaced by the Moyal $\star$-products.

We can get the NC QED Lagrangian by using above recipe as
\begin{equation}
{\cal L}=\frac{1}{2}i(\bar{\psi}\star \gamma^\mu D_\mu\psi
-(D_\mu\bar{\psi})\star \gamma^\mu \psi)- m\bar{\psi}\star
\psi-\frac{1}{4}F_{\mu\nu}\star F^{\mu\nu}, \label{NCL}
\end{equation}
where
$D_\mu\psi=\partial_\mu\psi-ieA_\mu\star\psi$,$~~(D_\mu\bar{\psi})=\partial_\mu\bar{\psi}+ie\bar{\psi}\star
A_\mu$, and $~~ F_{\mu\nu}=\partial_{\mu} A_{\nu}-\partial_{\nu}
A_{\mu}-ie(A_{\mu}\star A_{\nu}-A_{\nu}\star A_{\mu})$. This
Lagrangian can be used to obtain the Feynman rules for perturbation
calculations. The propagators for the free fermions and gauge fields
of NC QED are the same as that in the case of ordinary QED. But for
interaction terms, every interaction vertex has a phase factor
$exp(i p_1^{\mu}\theta_{\mu\nu}p_2^{\nu}/2)$ correction depending on
in and out 4-momentum. Moreover, there are cubic and quadric
interaction vertices for the gauge field besides the usual vertices
found in ordinary QED. The matrix $\theta_{\mu\nu}$ can be
decomposed into two independent parts \cite{Hewett01,Kamoshita}:
electric-like components $\bf{\theta}_E$ $=$
$(\theta_{01},\theta_{02},\theta_{03})$ and magnetic-like components
$\bf{\theta}_B$ $=$ $(\theta_{23},\theta_{31},\theta_{12})$.
$\bf{\theta}_E$ and $\bf{\theta}_B$ can be considered as 3-vectors
which define two unique directions in space. The characteristic
properties of NC QED may thus be observed as direction-dependent
deviations from the predictions of QED.

In this paper, we adopt the spirit of Ref.\cite{Hewett01} to
consider the possibility that NC scale energy $\Lambda_{NC}$ is
near the \tev~ scale, and study the scattering process
$e^+e^-\to\mu^+\mu^-\gamma$ in NC QED to establish the relation
for NC correction of scattering section versus collision energy,
and then to establish the relation between NC scale $\Lambda_{NC}$
and the optimal collision energy. Finally, the experimental
methods to determine the NC scale energy $\Lambda_{NC}$ and to
improve the precision of determining NC effects will be discussed,
because this process is an ${\cal O}(\alpha^3) $  NC QED process,
high precision tests has to be necessary.

\section{NC QED SCATTERING AMPLITUDES}

\begin{figure}
\includegraphics[scale=.6]{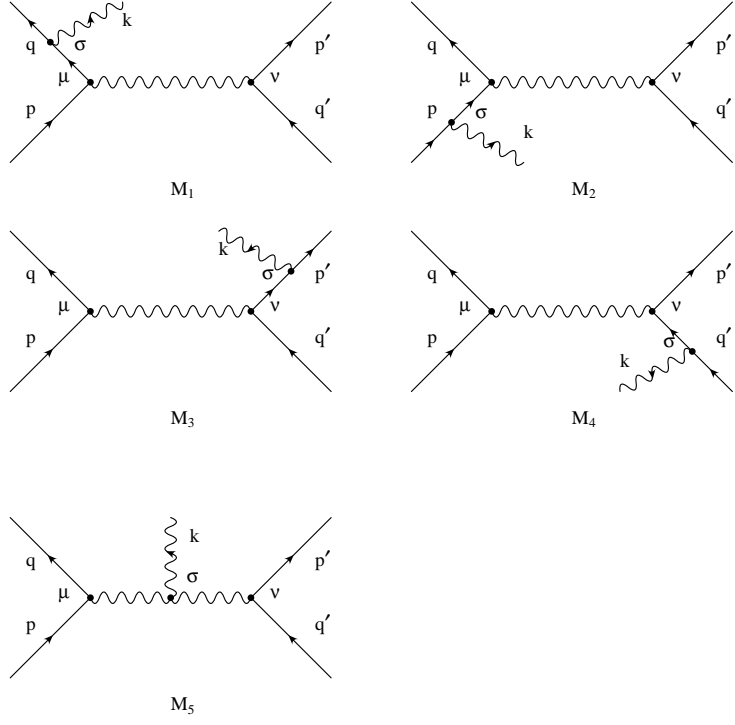}
 \caption{Feynman graph of
 scattering process $e^+e^-
\to\mu^+\mu^-\gamma$ } \label{Fig1}
\end{figure}

Shown in Fig.\ref{Fig1} are the five Feynman diagrams, which are
the lowest tree level contributions to the process.

Comparing with ordinary QED, we have one more Feynman graph, namely
Fig. $M_5$. This is due to the coupling of three photons in
noncommutative space-time. Using  the Feynman rules in
Ref.\cite{Sheikh-Jabbari99} and the Dirac equation, we get the
corresponding scattering amplitude as following \be
\begin{array}{ll} M_1=&\frac{ig^3 \varepsilon^*_\sigma(k)}{(p'+q')^2
(2 q\cdot k)}\bar{v}^{r_1}(q) \gamma^\sigma (-\qslash+\kslash+m)
\gamma^\mu u^{r_2}(p) \bar{u}^{s_1}(p')\gamma_\mu v^{s_2}(q')\\\\&
e^{i(p+q) \theta (-q+k)/2-iq' \theta p'/2},\end{array} \ee
\be
\begin{array}{ll} M_2=&\frac{ig^3 \varepsilon^*_\sigma(k)}{(p'+q')^2
(2 p\cdot k)}\bar{v}^{r_1}(q) \gamma^\mu (\pslash-\kslash+m)
\gamma^\sigma u^{r_2}(p) \bar{u}^{s_1}(p')\gamma_\mu
v^{s_2}(q')\\\\& e^{i(p+q) \theta (p-k)/2-iq' \theta p'/2},
\end{array}\ee
\be  \begin{array}{ll} M_3=&\frac{-ig^3
\varepsilon^*_\sigma(k)}{(p+q)^2 (2 p'\cdot k)}\bar{v}^{r_1}(q)
\gamma_\mu u^{r_2}(p) \bar{u}^{s_1}(p')\gamma^\sigma
(\pslash\,'+\kslash+\mu) \gamma^\mu v^{s_2}(q')\\\\& e^{i(p'+k)
\theta (p'+q')/2-ip \theta q/2}, \end{array}\ee
\be
\begin{array}{ll} M_4=&\frac{-ig^3 \varepsilon^*_\sigma(k)}{(p+q)^2
(2 q'\cdot k)}\bar{v}^{r_1}(q) \gamma_\mu u^{r_2}(p)
\bar{u}^{s_1}(p')\gamma^\mu (-\qslash\,'-\kslash+\mu) \gamma^\sigma
v^{s_2}(q')
\\\\&e^{i(-q'-k) \theta (p'+q')/2-ip \theta q/2}, \end{array}\ee
\be \begin{array}{ll} M_5=&\frac{-2g^3 \sin(k \theta (p+q)/2)
\varepsilon^*_\sigma(k)}{(p+q)^2 (p'+q')^2}\bar{v}^{r_1}(q)
\gamma^\mu u^{r_2}(p)\left[(-p-q-p'-q')^\sigma g_{\mu\nu}
+\left(p'\right.\right.
\\ \\& \left.\left.+q'-k\right)_\mu g^\sigma_\mu +(k+p+q)_\nu g^\sigma_\mu \right]
\bar{u}^{s_1}(p')\gamma^\nu
 v^{s_2}(q') e^{-i(p \theta q+q' \theta p')
/2},\end{array} \ee here $p$, $q$, $p'$, $q'$ and $k $ are four
momentums of electron, positron, muon, anti-muon and photon,
respectively; $r_1$, $r_2$, $s_1$, $s_2$ are spin indices of
corresponding particle, respectively; $m$, $\mu$ are masses of the
electron and the muon, respectively.

\section{SCATTERING CROSS SECTION}

The  cross section for process $e^+e^-\to\mu^+\mu^-\gamma $ is \be
\sigma=\frac{1}{16 w^2 v}\frac{1}{(2\pi)^5}\int|M(p,q\to p',q',k)|^2
(2\pi)^4 \delta^{(4)}(p+q-p'-q'-k)\frac{d\mathbf{p'}}{
p'_0}\frac{d\mathbf{q'}}{ q'_0}\frac{d\mathbf{k}}{ k_0},\ee
 where
 $p'_0$, $q'_0$, $ k_0$ are energies of
outgoing particles respectively, $v$ is the velocity of incident
electron, which is given by $v=(1-4m^2/w^2)^{1/2}$. In the
center-of-mass system C $(\mathbf{p}+\mathbf{q}=0)$, the total
energy of colliding electron-positron beams is $w^2=(p+q)^2=4E^2$,
where $E$ is the energy of each of the initial particles, the photon
energy $k_0$ may vary from a lower limit $\epsilon $ to an upper
limit
 $(w^2-4\mu^2)/2w$ which is determined by the kinematics. The limit
$\epsilon$ will in practice be identified with the soft-photon
cut-off.

\begin{figure}
\includegraphics[scale=.6]{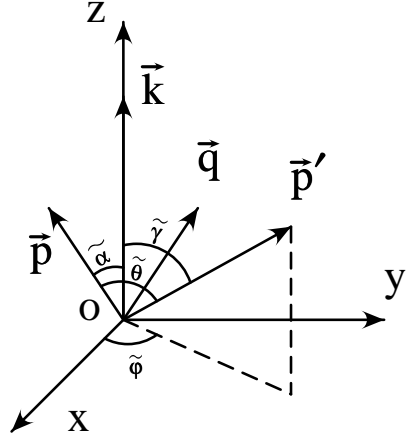}
\caption{The $C'$-system in which the center-of-mass of muons is at
rest} \label{Fig2}
\end{figure}

We integrate the cross-section $d\sigma$ over $ d\mathbf{p'}$ and
$ d\mathbf{q'}$ in the system $C'$ in which the center-of-mass of
the muons is at rest so that
$\mathbf{p'}+\mathbf{q'}=\mathbf{p}+\mathbf{q}-\mathbf{k}=\mathbf{0}$.
The system $C'$ is determined when the vector $\mathbf{k}$ in the
system $C$ is fixed. Its z-axis is along the orientation of
$\mathbf{k}$, the vectors $\mathbf{p}$, $\bf{q}$, $\bf{k}$ are
located in the plane $xoz$, the angles between $\bf{p'}$ and
$\bf{p}$ or $\bf{k}$ are $\tilde{\theta}$ or $\tilde{\gamma}$, the
azimuth angle of line $\bf{p'}$ is $\tilde{\varphi}$, the angle
between vector $\bf{p}$ and $\bf{k}$ is $\tilde{\alpha}$ (see
Fig.~\ref{Fig2}). The integration with respect to $\bf{p'}$ and
$\bf{q'}$  can be converted into integration with respect to
$\tilde{\gamma}$, $\tilde{\varphi}$.
Then differential cross section is, \be
\frac{d\sigma}{d\mathbf{k}}=\frac{\tilde{v}}{32 w^2 v (2\pi)^5 k_0}
\left.\int\limits_{-1}^1
d\cos\tilde{\gamma}\int\limits_0^{2\pi}d\tilde{\varphi}|M(p,q\to
p',q',k)|^2 \right|_{p+q-p'-q'-k=0},\ee here
$\tilde{v}=(1-4\mu^2/(p'+q')^2)^{1/2}$ is the velocity of muon.

In the C-system, the total cross-section $\sigma$, for the
emission of photons whose energy $k_0$ varies from $\epsilon $ to
$k_{0max}$, is given by \be
\sigma=\int\limits_{\epsilon}^{k_{0max}}\frac{d\sigma}{dk_0}dk_0
=\int\limits_{\epsilon}^{k_{0max}}dk_0\int\frac{d\sigma}{d\bf{k}}d\Omega(\bf{k}).\ee

\section{NC CORRECTION AND NUMERICAL ANALYSIS}

After all the integration, we obtain the following expression of
cross-section, \be \sigma=\sigma_{QED}+(\Delta\sigma)_{NC}.\ee The
first term is the usual QED cross section\be
\sigma_{QED}=\int\limits_{\epsilon}^{k_{0max}} dk_0
\left(\frac{\left((1-v^2)w^2+2\rho^2\right)\gamma_0+4{k_0}^2
\gamma_1}{\left(3-v^2\right)\rho^2
k_0}+\frac{\left((1-\tilde{v}^2)\frac{\rho^4}{w^2}+2\rho^2\right)\gamma_0'+4{k_0}^2
\gamma_1'}{\left(3-\tilde{v}^2\right)w^2 k_0}\right)\sigma_0; \ee
with \be \sigma_0=\frac{1}{3}\frac{\pi\alpha^2}{
w^2}\frac{\tilde{v}}{v}(3-v^2)(3-\tilde{v}^2),\ee \be
\begin{array}{l}\gamma_0=\frac{2\alpha}{\pi}\left(-1+\frac{1+v^2
}{2v}\ln\frac{1+v}{1-v}\right), \\\\\gamma_1=\frac{2\alpha}{
\pi}\left(-1+\frac{1}{v}\ln\frac{1+v}{1-v}\right),
\\\\  {\gamma_0}'=\frac{2\alpha}{\pi}\left(-1+\frac{1+\tilde{v}^2
}{2\tilde{v}}\ln\frac{1+\tilde{v}}{1-\tilde{v}}\right),
\\\\  {\gamma_1}'=\frac{2\alpha}{\pi}\left(-1+\frac{1
}{\tilde{v}}\ln\frac{1+\tilde{v}}{1-\tilde{v}}\right),
\end{array}\ee
here $\alpha$ is the fine structure constant.

The second term is the NC correction defined as the difference
between NC QED cross section and commutative QED cross section
 \be
(\Delta\sigma)_{NC}=2 \alpha^3\int\limits_{-1}^1dz\int\limits_0^{2
\pi}d\phi\int\limits_{\epsilon}^{k_{0max}}dk_0\frac{\tilde{v}}{
v}\frac{k_0}{w^4\rho^2}\sin^2\frac{k \theta (p+q)}{2 }
f(w,k_0,z),\ee where \be \sin^2\frac{k \theta (p+q)}{ 2
}=\sin^2[\frac{-1}{2\Lambda_{NC}^2}wk_0(c_{01}
\sin\gamma\cos\phi+c_{02}\sin\gamma\sin\phi+c_{03}\cos\gamma)],
\label{cmatrix}\ee and $\gamma$, $\phi$ are the polar angle and
azimuth angle of photon in C-system, and \be
\begin{array}{l}
f(w,k_0,z)=\frac{1}{8\pi w^2 k_0^2}\left[- \frac{16w^4 k_0^2 x}{3(
1 - vz) } - 8w^2{\rho }^4 -
    4Bw^2{\rho }^2\left( w^2 + {\rho }^2 \right)\right. \\\\ +
    y\left( w^6 - 2w^4{\rho }^2 +5w^2{\rho }^4 - 2{\rho }^6 \right) +
       2Bw^2{\rho }^2\left( w^2 + {\rho }^2 \right)\left( x+y \right)  +
    \frac{4}{3}x\left( w^2 + {\rho }^2 \right) \left( w^4 + {\rho }^4 \right)  \\\\
     + \left( 1 - vz \right) {\rho }^2 \left( 4Bw^2{\rho }^2 - 2w^2\left( w^2 - 3{\rho }^2 \right)
     - \frac{2}{3}x{\rho }^2\left(3Bw^2 + 3w^2 + {\rho }^2  \right)  \right)  \\\\ +
    xy\left( \frac{4w^2\left( w^4 - {\rho }^4 \right) }{3\left( 1 - v^2z^2 \right) }-3Bw^2{\rho }^4 -
       \frac{w^8 + 2w^6{\rho }^2 - 2w^4{\rho }^4 + 12w^2{\rho }^6 - {\rho }^8}{3{\rho }^2}
       +w^2{\rho }^2\frac{3B\left( w^2 + {\rho }^2 \right)  + 4\left( w^2 + 2{\rho }^2 \right) }
        {3(1 + vz)} \right)  \\\\
        + \frac{24w^2{\rho }^4 + 12Bw^2{\rho }^2\left( w^2 + {\rho }^2 \right)
        - 6y\left( 2w^2{\rho }^4 + Bw^2{\rho }^2\left( w^2 + {\rho }^2 \right)  \right)
       - x\left( 6Bw^2{\rho }^2\left( w^2 + {\rho }^2 \right)  + 4w^2\left( w^4 + 3{\rho }^4 \right)
          \right) }{3(1 + vz)} \\\\
         + \left( 1 - v^2z^2 \right)
\left( -w^6 + 4w^4{\rho }^2 - 11w^2{\rho }^4 + 10{\rho }^6 +
       2B{\rho }^2\left( w^4 - 5w^2{\rho }^2 + 2{\rho }^4 \right)
      \right)\\\\ +\left.
       x  \left( 1 - v^2z^2 \right)\left( - B{\rho }^2\left( 2w^4 - 7w^2{\rho }^2 + 2{\rho }^4 \right)    +
          \frac{5w^8 - 27w^6{\rho }^2 + 57w^4{\rho }^4 - 13w^2{\rho }^6 - 6{\rho }^8}{6w^2}   \right)
    \right].
\end{array}
\ee Here  $x=3-\tilde{v}^2$, $y=3-v^2$,
 $z=\cos\gamma$, $B=\frac{1}{\tilde{v}} \ln \frac{1 - {\tilde{v}}}{1 +
       {\tilde{v}}}$, and $\rho^2=w^2-2\,w\,k_0$.

When $\theta \to 0$, the total cross-section reduces to the ordinary
QED result.

\subsection{THE MATRIX $\theta_{\mu\nu}$ IN THE LABORATORY SYSTEM}

From the eq.(\ref{cmatrix}), we find that the process $e^+e^-\to
\mu^+\mu^- \gamma$ is sensitive only to $\bf{\theta}_E$, but not
to $\bf{\theta}_B$ at tree level. Here $\bf{\theta}_E$ is defined
as \be {\bf\theta_E}=\frac{1}{\Lambda_{NC}^2}{\bf
c_E}=\frac{1}{\Lambda_{NC}^2}(c_{01},c_{02},c_{03}),\ee where
$\Lambda_{NC}=1/\sqrt{|\bf{\theta}_E|}$ and $c_{0i}$
 are components of the unit vector $\bf{c}_E$ pointing to the
 unique direction in the primary coordinate system.
 In the general case, $\sin^2\frac{k \theta (p+q)}{2}$ depends
 not only on the photon polar angle $\gamma$, but also on
 the azimuthal angle $\phi$. This is a signature of the anisotropy
 of space-time which is inherent in noncommutative geometry.
 In the special case $(c_{01}=c_{02}=0)$, the $\bf{\theta}_E$ is independence of $\phi$.

It would be natural to assume that the unique direction $\bf{c}_E$
is fixed to some larger structure in space, e.g., the frame of the
 cosmic microwave background which could be taken as primary frame. In the Lab coordinate system  on Earth,
$\bf{c}_E$  will change as the Earth rotates and as the
 orientation of the Earth's rotation axis changes due to the
 movement of the galaxy or the solar system with respect to the
 primary frame. We assume that the latter movement is sufficiently
 slow, the rotation of the Earth is only the relevant motion in the time scale of the experiment.

 In the primary frame, the unit vector $\bf{c}_E^0$ is specified
 by two parameters, the polar angle $\eta$ and the azimuthal angle
 $\xi$\cite{Kamoshita}:
 \be \bf{c}_E^0=\left(\begin{array}{c} s_\eta c_\xi \\s_\eta
 s_\xi\\ c_\eta \end{array}\right),\ee
 where $s_\eta\equiv\sin\eta$, $c_\xi\equiv\cos\xi$ and so on.

 Elements of the vector $\bf{c}_E$ in the local coordinate system
 (x, y, z) are obtained by successive rotations of the coordinate
 axes:
 \be\begin{array}{ll}\bf{c}_E & =R_y(\beta)R_z(-\pi/2)R_y(-\delta)R_z(\zeta)\cdot \bf{c}_E^0
 \\\\
& =\left(\begin{array}{c} s_\beta s_\delta \\ c_\delta
 \\-c_\beta s_\delta \end{array}\right) s_\eta
 \cos(\zeta-\xi)+\left(\begin{array}{c}c_\beta \\ 0 \\ s_\beta
 \end{array}\right) s_\eta \sin(\zeta-\xi)+\left(\begin{array}{c}-s_\beta c_\delta \\ s_\delta \\
 c_\beta c_\delta
 \end{array}\right) c_\eta,\label{anglel}\end{array}\ee
 here the angle $\delta$ is the latitude of the experiment
 location, $\beta$ is the angle between the z-axis and the
 direction of north, $\zeta$ is a time-dependent azimuthal angle of
the experiment site, given by\cite{Abbiendi03}:
 \be\zeta(t)=\omega t,\label{timel}\ee
 where $\omega=2\pi/T_{sd}$ with sidereal day $T_{sd}$.

 So the $c_{0i}$  will vary with experiment site, i.e., with values of $\beta$, $\delta$ and
 $\eta$, and the variation of $c_{0i}$  with time is decided by $\zeta-\xi$.
 When $\beta=\delta=\eta=0$, we get $c_{03}=1$, and $c_{01}=c_{02}=0$.

 \subsection{ NUMERICAL ANALYSIS}

\begin{figure}
\includegraphics[scale=.6]{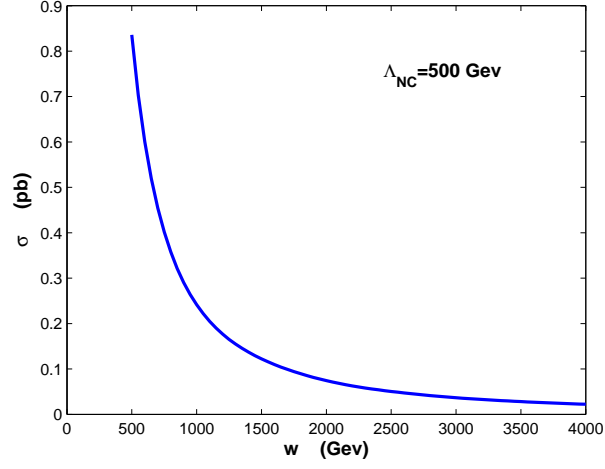}
\caption{The relation between $\sigma$ and $w $ for
$e^+e^-\to\mu^+\mu^-\gamma$} \label{Fig3}
\end{figure}

\begin{figure}
\includegraphics[scale=.6]{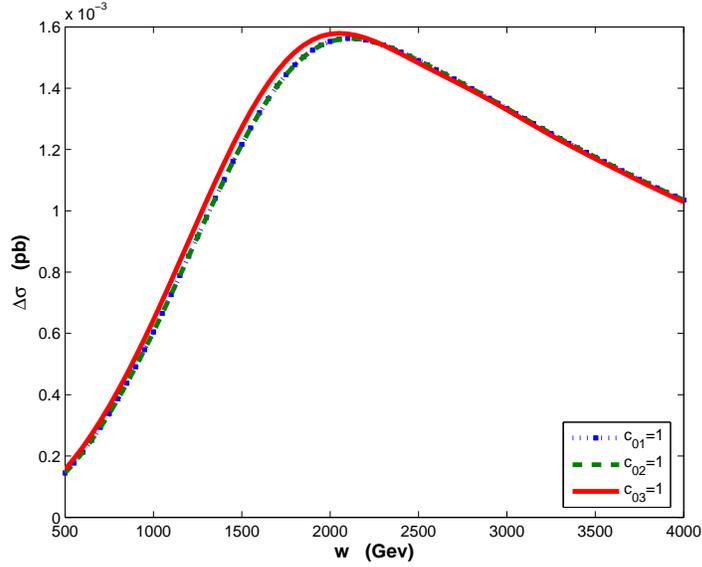}
\caption{The relation between ${(\Delta\sigma)}_{NC}$ and $w$ for
$e^+e^-\to\mu^+\mu^-\gamma$} \label{Fig4}
\end{figure}

Now we consider NC correction to the total cross-section for the
process $e^+e^-\to\mu^+\mu^-\gamma$ by numerical integration. The
energy of the photon varies from $\epsilon=0.01\mu$ to
$k_{0max}=2w/5$ in our case. The relation curve between
cross-section $\sigma$ and collision energy $w$ is shown in
Fig.~\ref{Fig3}. The cross-section of this process decreases with
collision energy $w $, which is similar to ordinary QED.
The NC correction for the cross-section varies with $w$ is shown in
Fig.~\ref{Fig4}, where $\Lambda_{NC}=500\gev$.

From Fig.~\ref{Fig4}, it is found that the differences among the
curves for $c_{01}=1$, $c_{02}=1$ and $c_{03}=1$ are small. In
fact, the curves of ${(\Delta\sigma)}_{NC}$ with $c_{01}=1$ and
$c_{02}=1$ are almost identical. We will only discuss the case
with $c_{03}=1$ in the following analysis if there is no
indication.

\begin{figure}
\includegraphics[scale=.6]{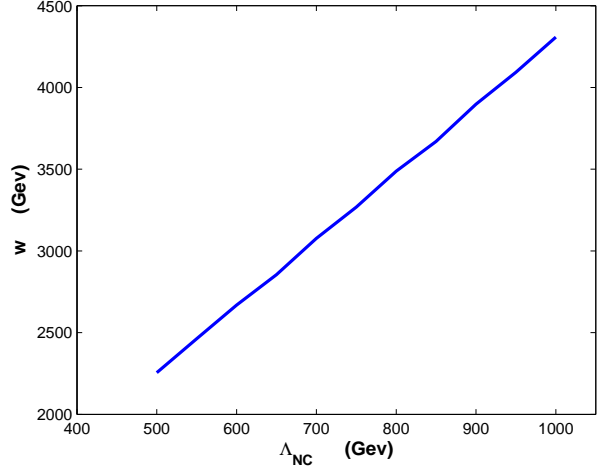}
\caption{The optimal collision energy $w_o$ vs  NC scale energy
$\Lambda_{NC}$ for $e^+e^-\to\mu^+\mu^-\gamma$} \label{linear}
\end{figure}

${(\Delta\sigma)}_{NC}$ does not increase monotonously with $w$,
but there exists a kurtosis distribution. For
$\Lambda_{NC}=500\gev$, a maximum appears when the energy of
collision particles $ w =2258\gev$. It implies that there is an
optimal collision energy to observe NC effect in this process.
However, we do not know the exact value of NC scale energy
$\Lambda_{NC}$. We have to determine it at first. To do this, it
is useful to establish the relation between NC scale energy and
optimal collision energy $w_o$, which is shown as in
Fig.\ref{linear}.
Numerically, we get \be
w_o=229.263+4.067\Lambda_{NC}\;\;\;\;(\gev).\label{linear1} \ee
Roughly, the optimal collision energy is about four times of NC
scale energy. In principle, the NC effects could be detected when
the particle energy is higher than $\Lambda_{NC}$, and the NC
scale energy $\Lambda_{NC}$ is nothing but the characteristic
energy at which the difference between QFT and NC QFT emerges.
However, it is very difficult to determine directly such a
characteristic energy, because the difference increases gradually,
so that we do not know where is the start point of the difference.
Determination of an optimal collision energy at which NC
correction of scattering cross section arrives at the maximum is
much easier, because determination of a highest point of curve is
much easier than determination of an inflexion of curves in
mathematics. If the optimal collision energy $w_o$ for NC effect
is determined in the next generation colliding experiment by
increasing gradually the energy of collision particles, then the
NC QED scale energy $\Lambda_{NC}$ can be determined by relation
eq.(\ref{linear1}). This is an indirect but effective way. It is
interesting to note that this linear relation also arise in
M\"{o}ller scattering and Bhabha scattering \cite{sheng}.

Because this process is higher order in $\alpha$ ( ${\cal
O}(\alpha^3) $ ) comparing with other $e^+e^-\to l^+ l^-$
processes, its value of ${(\Delta\sigma)}_{NC}$ is less than one
percent of others. The higher precision tests, thus, has to be
necessary, which means the luminosity of electron-positron beam
should be increased (e.g. $ 10^{35}cm^{-2}s^{-1} $) so that this
effect can be determined. In order to increase the percentage of
NC correction to the cross section, another method is to use
directly the anisotropic property of space-time, which is inherent
characteristic of noncommutative geometry. This anisotropy will be
eliminated partly after the integration respect to both polar
angle and azimuthal angle for total scattering cross section.
Measuring the difference of the differential cross sections at
different polar angles not only can keep this inherent anisotropy,
but also can reduce the systematic error of measurements.

\begin{figure}
\includegraphics[scale=.6]{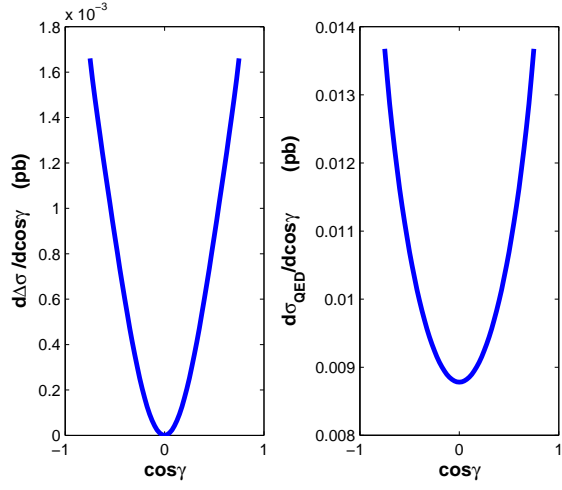}
\caption{The differential cross-section and its NC correction vs
$\cos\gamma$}
 \label{differential}
\end{figure}

When collision energy $w=2258 Gev$ and $\Lambda_{NC}=500 Gev$, the
differential cross-section for commuting QED(right) and their NC
correction(left) varying with $\cos\gamma$ are shown as in
Fig.\ref{differential}. The difference of the differential cross
section for QED and its NC correction  between $\gamma=\pi/4$ and
$\gamma=\pi/2$ is  0.0042pb and 0.0015pb respectively. Their ratio
is $35.7\%$, which is much larger than the ratio between  total
cross section and  its NC correction. It should be noted that the
observation angles $\gamma$ should be not less than $\pi/4$ in
order to obtain a larger ratio, because the differential cross
section for commuting QED has a remarkably nonlinear increase with
$\cos\gamma$ in contrast to slowly increase for its NC correction
when $\cos\gamma >0.7$ or  $\gamma < \pi/4$. The NC correction of
the differential cross section is also small although this ratio
is large, so  the NC effects can be detected only when the
luminosity of electron-positron beam is large enough.

\section{CONCLUSION AND DISCUSSION}

  The noncommutative corrections to the cross section
for process $e^+e^-\to\mu^+\mu^-\gamma$ is studied and three new
results are obtained. First, the higher collision energy in
$e^+e^-\to\mu^+\mu^-\gamma$ does not always provide a better
avenue to explore noncommutative effect. But there is an optimal
collision energy to get the greatest noncommutative correction to
total scattering section. In order to explore the noncommutative
effects in the next generation international linear collider, we
shall choose the optimal collision energy as the running energy
with enough luminosity of electron-positron beam. This conclusion
is reasonable, because the noncommutative effects do not more
readily appear at low collision energy ( the collision energy must
be higher than the NC scale energy $\Lambda_{NC}$ ), and the NC
correction will decrease due to the total scattering section
decrease with collision energy increase.

Second, there is a linear relation between the optimal collision
energy and the NC QED scale energy $\Lambda_{NC}$ given by
eq.(\ref{linear1}). If the optimal collision energy is determined
by increasing gradually the energy of collision particles from
hundreds of $\gev$ to several $\tev$ in the next generation linear
collider, then the NC QED scale energy can be determined from the
given linear relation in an indirect but effective way.

Third, there is an experimental method to improve the precision of
determining NC effects by measuring the difference of the
differential cross sections at different polar angle; which not
only can keep the inherent anisotropy, but also can reduce the
system error of measurements. It is significant to explore NC
effects for the process $e^+e^-\to\mu^+\mu^-\gamma$, because this
process is one of the higher precision experiments to test
ordinary QED.

In presenting these results, the Z-boson exchange contribution has
not been included. If Z-boson and photon have the same vertex
structure assumed as in Ref.\cite{Hewett01}, we can argue that the
linear relation between optimal collision energy and NC QED scale
energy$\Lambda_{NC}$ will hold, and we only need to modify the
proportionality constant to account for the Z-boson exchange
contribution.

It is interesting to note that, because of the anisotropy of
space-time, the NC effects will be different at different site and
different time for the same experiment. The dependence of NC
corrections to differential cross section on longitude and latitude
of experiment site and time is much stronger than that for total
cross section. It is possible  to determine the NC effects by
observing the change of differential cross section with time, but we
have to choose proper experiment site.

\nn{\bf Acknowledgments:}

The authors would like to thank referee's comments and
suggestions, and M.X. Luo and H.B. Yu for their useful
discussions, also thank the hospitality and support for the Abdus
Salam International Centre for Theoretical Physics, Trieste,
Italy, where this work is completed mainly. This work is also
supported in part by the funds from NSFC under Grant No.90303003.


\end{document}